# Room Temperature Spin Filtering and Quantum Transport with Transition Metal-Doped Silicon Quantum Dot

Hemant Arora[1] and Arup Samanta*[1,2]

[1]Quantum/Nano Science and Technology Laboratory, Department of Physics, Indian Institute of Technology Roorkee, Roorkee-247667, Uttarakhand, India
[2]Centre of Nanotechnology, Indian Institute of Technology Roorkee, Roorkee-247667, Uttarakhand, India
Corresponding author: arup.samanta@ph.iitr.ac.in

Spin filtering is a fundamental operation in spintronics, enabling the generation and detection of spin-polarized carriers. Here, we proposed and theoretically demonstrated that a 3d transition metal (TM) doped silicon quantum dot (SiQD) is a suitable candidate for spin filter device at room temperature. Using density functional theory (DFT), we investigate the structure, electronic properties, and magnetic behavior of TM-SiQD. Our calculations demonstrate that Mn-doped SiQD exhibits the highest stability. The designed spin-filter device using Mn-doped SiQD shows a spin-filtering efficiency of 99.9% at 300K electrode temperature along with very high conductance. This remarkable efficiency positions it as a promising candidate for room-temperature spintronic devices.

The generation and manipulation of spin-dependent transport properties hold significant importance in diverse fields, including spintronics, quantum information processing, and magnetic memory devices.[1–3] The challenges in the above-mentioned fields are to generate and control stable spin-charge carriers effectively. To tackle these challenges, spin filters have emerged as the most promising and versatile devices. A spin filter is a device that selectively allows the passage of spin-up and spin-down electrons and controls their transmission or reflection.[4,5] Spin filters can be implemented utilizing a wide range of materials and structures, including magnetic oxides, heavy metals, topological insulators with high spin-orbit coupling, and semiconductor structures.[6–10] Additionally, numerous researchers reported magnetic semiconductor or insulator thin films and 2D van der Waals (vdW) hetero-structures have also demonstrated the capability to achieve high spin filtering.[11-14] Among the above-mentioned, semiconductor nanostructures (e.g., quantum dots, quantum wells, and nanowires) have gained the most attention due to their adjustable electric and chemical properties. Efficient spin filters can be designed by precise controlling of electronic properties of the semiconductor nanostructure using confinement dimensions, material composition, and suitable doping.[15-18]

The widely explored technique for controlling the properties of semiconductors involves impurity doping, known as diluted magnetic semiconductors (DMSs).[19,20] Significant research efforts have been devoted to doping bulk semiconductors with magnetic ions, aiming to introduce extraordinary phenomena such as the giant Zeeman effect, Faraday rotation, and magnetic polaron effects. The presence of magnetic ions in DMS introduces a localized magnetic moment, which can couple with the electron spins in the material. This coupling, known as exchange interaction, leads to the spin-dependent behavior of charge carriers, allowing for spin filtering without using magnetic electrodes. Till now a lot of experimental and theoretical studies have been done for bulk material-based DMSs.[21-24] Among all of them, the Group IV-based DMSs have gained special interest due to their potential to get high Curie temperature ($T_c$).[25] In group IV, silicon has grabbed much attention over all the semiconductor family because of its high abundance and well-established CMOS-compatible device fabrication technology. Silicon, being an ideal host of spin-based technology, benefits from its low atomic mass and crystal inversion symmetry, resulting in minimal spin-orbit interactions. These characteristics contribute to efficient spin transport in silicon-based systems.[26] Considering such a promising advantage, recent articles have reported that bismuthene-silicon nanostructures, featuring cowrie shell-like geometries, could potentially function as efficient spin filtering devices.[8-10] However, it is important to note that, as of now, no experimental evidence has been provided to support this approach, primarily due to the complexity of material processing for such structures. Apart from this, the dopant-based silicon nanostructures have gained more attention because of the experimental feasiblilty[27,28] and scope in various applications such as single electron transistors,[29-34] spin qubits,[35-37] and spintronics.[38] There are a few reports available on spin transport in phosphorous-doped silicon nanowires at very low temperatures.[39,40] In contrast, TM-doped silicon nanostructures exhibit greater potential, given their ability to display ferromagnetism above room temperature.[27,28,41] This makes the TM-doped silicon nanostructures particularly promising for room temperature spintronic applications. Despite this potential, the study of spin transport behavior in silicon nanostructures is yet to be explored.

In this letter, keeping in view the experimental feasibility of the SiQD structures, we aim to advance the comprehension of TM impurity-induced magnetism in substitutional doped hydrogenated silicon quantum dot (H-SiQD). Further, this system is used to fabricate a room-temperature spin-filtering device. For this, we provide a detailed and systematic comparative study based on first principles calculation for the stability, electronic, and magnetic properties of TM atoms doped H-SiQD and present the transmission characteristics for spin filter devices.

The structure of a bare silicon quantum dot (b-SiQD) with a diameter of approximately 1.6 nm is formed using VESTA software.[42] The b-SiQD is obtained by taking bulk silicon atoms within a 1.6 nm radius sphere as shown in Fig. 1(a). The surface dangling bonds of b-SiQD are passivated by hydrogen atoms as shown in Fig. 1(b). Experimentally, SiQD having a diameter as small as 1 nm is formed.[43]

Before considering substitutional TM doping in SiQD, the band structures of b-SiQD and H-SiQD were studied using the PBE approach as shown in Fig. 1(d and e). The band structure of b-SiQD showed metallic behavior due to the presence of an unsaturated dangling bond on the surface and similar behavior is observed by the HSE approach (Fig. S1(a)).[44] After the termination of these dangling bonds by H atoms, H-SiQD showed the semiconducting behavior with a 2.65 eV bandgap. In addition, to overcome the GGA-PBE error, we employed the band structure calculation using the HSE approach revealing a band gap of 3.5 eV shown in Fig. S1(b), which is consistent with previously reported work.[45] This result indicates that our simulations can well capture the electronic structures of the host semiconductors and give us confidence for further investigating the electronic and magnetic properties of the corresponding TM:H-SiQD.

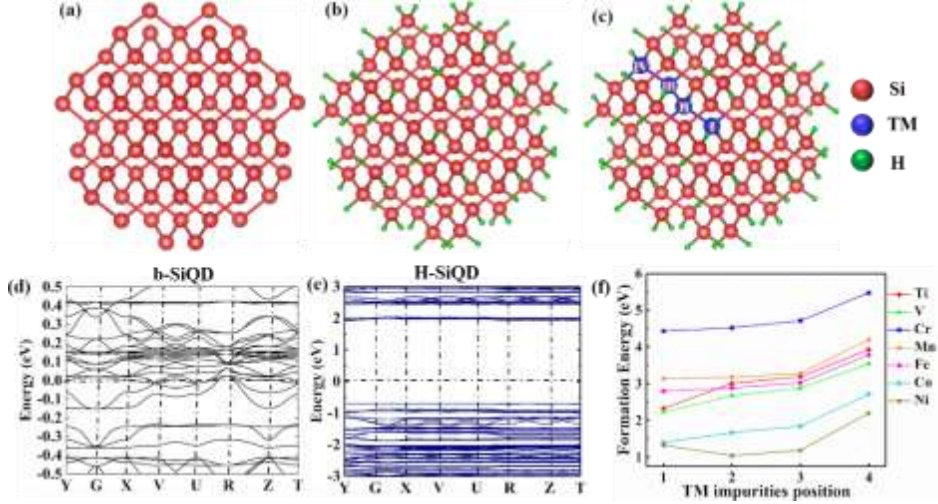

FIG.1. Atomic structure of (a) b-SiQD (Si$_{87}$), (b) H-SiQD (H$_{76}$Si$_{87}$), (c) TM atom doped at four different positions in H-SiQD. Band structures of (d) b-SiQD, (e) H-SiQD, and (f) The formation energies of 3d TM impurities at different positions in H-SiQD by PBE approach.

Further, we substituted 3d TM atoms at four different substitutional sites (I to IV) in H- SiQD (Fig. 1(c)) and found the most stable site by calculating formation energy using the Eq. (2)

$$E_f = E_{TM:H-SiQD} - E_{H-SiQD} - n\mu_{TM} + m\mu_{Si} \quad (2)$$

Where, $E_{TM:H\text{-}SiQD}$ represents the total energy of TM:H-SiQD, while $E_{H\text{-}SiQD}$ represents the total energy of H-SiQD. The energy of the TM and the silicon atom in the same lattice parameter as H-SiQD are referred to as $\mu_{TM}$ and $\mu_{Si}$, respectively. The larger value of $E_f$ indicates the requirement of higher energy for such a kind of substitutional doping. We found Mn and Fe impurities exhibit relatively stable behavior, with minimal changes (around 0.1 eV) from site I to III, indicating their stability in H-SiQD as shown in Fig. 1(f).[46] Further, we compared formation energies of central site-substituted TM:H-SiQD using both PBE and HSE approaches calculated by using Eq. (2) revealing the similar trend as listed in Table 1. Sequentially, we analyzed the magnetic stability of TM:H-SiQD configurations using the expression below

$$\triangle E = E_T^{up} - E_T^{sp} \quad (3)$$

Where $E_T^{up}$ and $E_T^{sp}$ represent the total energy of spin unpolarised and spin-polarized configurations. When $\triangle E > 0$ ($\triangle E \leq 0$) the configuration possesses a stable spin-polarized (spin-unpolarized) state. Results presented in Table 1 demonstrate that V, Cr, and Mn-doped H-SiQDs exhibit more stable magnetic phases in both approaches. A notable discrepancy is observed for Fe:H-SiQD, displaying non-magnetic behavior in PBE and magnetic behavior in HSE. Additionally, Table 1 lists total magnetic moments ($\mu_{Total}$), which sum the magnetic moments of all atoms, and local magnetic moments ($\mu_{Local}$) of TM atoms at the central substitutional site. Here, V, Cr, and Mn indicate the spin polarization of Si atoms is opposite to the polarization of TM atoms, whereas Fe and Co exhibit the reverse.[41] The above calculations showed that the Mn and Fe-doped

TABLE 1. List of optimized bond length, Bond angles, formation energy ($E_F$), total magnetic moment ($\mu_{Total}$), local magnetic moment ($\mu_{Local}$), and magnetic stability energy ($\triangle E$) for TM:H-SiQD configurations.

| Dopant | | Ti | V | Cr | Mn | Fe | Co | Ni |
|---|---|---|---|---|---|---|---|---|
| TM-Si (Å) | | 2.54 | 2.42 | 2.37 | 2.28 | 2.36 | 2.25 | 2.26 |
| Bond angles (degree) | | 108.6-109.9 | 109.4-109.6 | 109.5-109.9 | 104.4-111.9 | 105.9-116.5 | 109.1-109.7 | 95.6-118.6 |
| $E_F$ (eV) | PBE | 2.33 | 2.23 | 4.44 | 3.14 | 2.8 | 1.4 | 1.33 |
| | HSE | 3.25 | 2.69 | 4.66 | 2.93 | 2.55 | 1.46 | 1.23 |
| $\mu_{Total}$ ($\mu_B$) | PBE | 0 | 1 | 2 | 3 | - | 1 | 0 |
| | HSE | 0 | 1 | 2 | 3 | 4 | 1 | 0 |
| $\mu_{Local}$ ($\mu_B$) | PBE | 0 | 1.108 | 3.737 | 3.290 | - | 0.623 | 0 |
| | HSE | 0 | 1.297 | 3.721 | 3.745 | 3.853 | 1.735 | 0 |
| $\triangle E$ (eV) | PBE | 0 | 0.25 | 3.19 | 0.31 | -0.18 | 0.05 | 0 |
| | HSE | 0 | 0.83 | 3.2 | 1.81 | 1.24 | 0.4 | 0 |

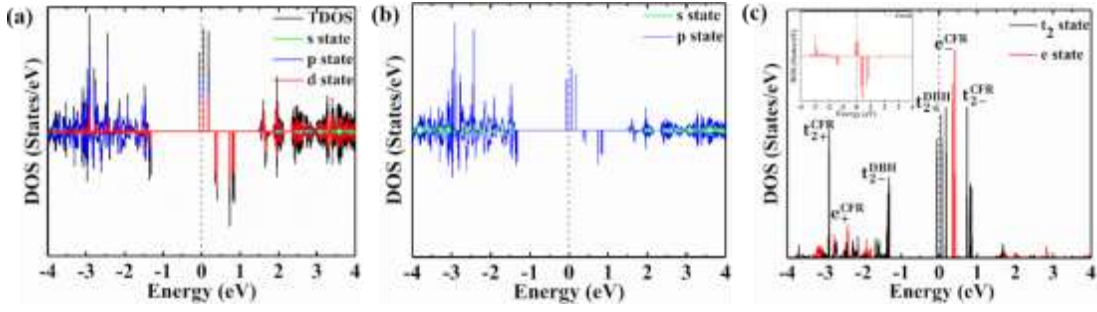

FIG. 2. Spin-up and spin-down (a) TDOS and orbital decomposed DOS, (b) orbital decomposed DOS of Si atoms, and (c) Mn atom PDOS, of Mn:H-SiQD calculated by PBE approach where the $t_2$ and e symmetry as well as spin (+ and -) have been indicated. The d orbital DOS of the Mn atom is given in the inset.

H-SiQD across various dopant sites, high magnetic stability, and significant magnetic moments. These traits indicate their potential for Si atom substitution and magnetism in the H-SiQD configuration.

For the complete evaluation of electronic properties and origin of magnetism, we determined the total density of states (TDOS), orbital decomposed density of states, and partial density of states (PDOS) for Mn and Fe-doped H-SiQD by PBE and HSE approaches. For Fe:H-SiQD, due to instability in the PBE approach, as we discussed above, the DOS is calculated by the HSE approach and presented in the supplementary file as Fig. S3. The DOS exhibited near-identical spin-up and spin-down states in the valance and conduction band, with additional states near the Fermi level indicating a flat band nature, influencing localized magnetic moments. The band structure calculation is shown in the supplementary file Fig. S2 with listed bandgap values in Table S1. The origin of magnetism in TM:H-SiQD is due to the strong p-d hybridization between Si 3p and TM 3d orbitals, evident in total DOS and orbital decomposed DOS of Si and TM as presented in Fig. 2 and Fig. S3. This hybridization leads to tetrahedral crystal field splitting, forming bonding and anti-bonding $t_2$ states and nonbonding e states. These hybridized states are evident in TM PDOS (Fig. 2(c) and S3(c and f)), distinguished by spin-up and spin-down states using orbital decomposed DOS of TM d states (inset of PDOS). In Mn:H-SiQD, $t_{2+}^{CFR}$ is lower, suggesting strong p-d coupling. On the other hand in Fe:H-SiQD, $t_{2+}^{CFR}$ is close to $e_+^{CFR}$, indicating lower stability than Mn:H-SiQD. In both Mn and Fe doped H-SiQD, the degeneracy of $t_{2+}^{DBH}$ states are broken due to variations in bond length and angle, contributing to the Jahn-Teller effect.[47-49] In Mn:H-SiQD, the substantial variation in bond length and angle, and in Fe:H-SiQD, the high variation in bond angles leads to a stretched tetrahedral geometry. In Mn:H-SiQD (shown in Fig. S2(a and c)), one $t_{2+}^{DBH}$ the state lies below and two states are above the Fermi level, indicating a shift of the Fermi level towards the energy level associated with the majority charge carriers. Similarly, in Fe:H-SiQD (shown in Fig. S2(e)), two spin-up states can be observed below the Fermi level due to the presence of two electrons in the $t_{2+}^{DBH}$ level. Based on the findings, it is suggested that Mn:H-SiQD has the potential to function as a spintronic device.

Following the comprehensive analysis of all properties associated with Mn:H-SiQD, we employed the DFT-NEGF approach to investigate the spin-dependent transport characteristics in the Au-Mn:H-SiQD-Au device, as illustrated in Fig. 3.[50] The transport characteristics have been measured by entering the spin-unpolarized electrons from the source terminal and collection of spin-polarized electrons at the drain terminal. We ultimately measured the transmission probabilities of spin-up ($T_{up}$) and spin-down ($T_{down}$) electrons at energy E. Subsequently, we determined the spin polarization (P) of the electrons exiting through the drain contact using the formula

$$P = \frac{T_{up}}{T_{up}+T_{down}} \quad (4)$$

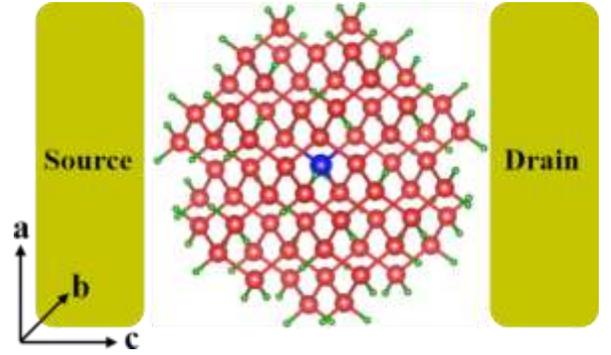

FIG. 3. A schematic of Mn:H-SiQD spin filter device

The concept of spin polarization, quantifying the alignment of electron spins, was employed to gain insights into the specific orientation of the electrons' magnetic properties. To delve deeper into the behavior of electrons in different spin states, we strategically manipulated the initial spin of the Mn atom, oriented along a-, b-, and c-directions under various bias voltages. By applying a spin-conserved tunneling method, we elucidated the transmission functions for spin-up and spin-down carriers, represented visually in red and blue, respectively (as depicted in Fig. 4). Our in-depth analysis revealed a fascinating phenomenon that the transmission peaks of both spin-up and spin-down carriers corresponded with DOS of the Mn-SiQD nanostructure. This complex alignment highlighted the presence of a spin-dependent resonant scattering mechanism commonly coined as Resonant scattering. Resonant scattering is a significant phenomenon that occurs when the energy of transmitted electrons precisely matches specific eigenvalues of the nanostructure's Hamiltonian. These eigenvalues correspond to energy eigenstates of the isolated nanostructure. As electrons traverse the nanostructure, they engage in partial scattering events, both forward and backward, through these eigenstates. This interaction results in distinctive peaks (resonances) or dips (anti-resonances) in the electron transmission probability concerning electron energy. These phenomena are essential in understanding the behavior of electrons in quantum systems and form the basis for advanced electronic devices. In our system, the presence of ferromagnetism induces a separation between the spin-up and spin-down states. Consequently, the electron transmission probabilities for spin-up and spin-down electrons differ, giving rise to spin filtering. This phenomenon occurs due to the distinct behavior of spin-up and spin-down electrons, leading to selective transmission and ultimately enabling spin-dependent filtering in our device. Further, we have also calculated the conductance of the structure using the Landauer's formula

$$G(E) = \frac{e^2}{h}(T_{up}(E) + T_{down}(E)) \quad (5)$$

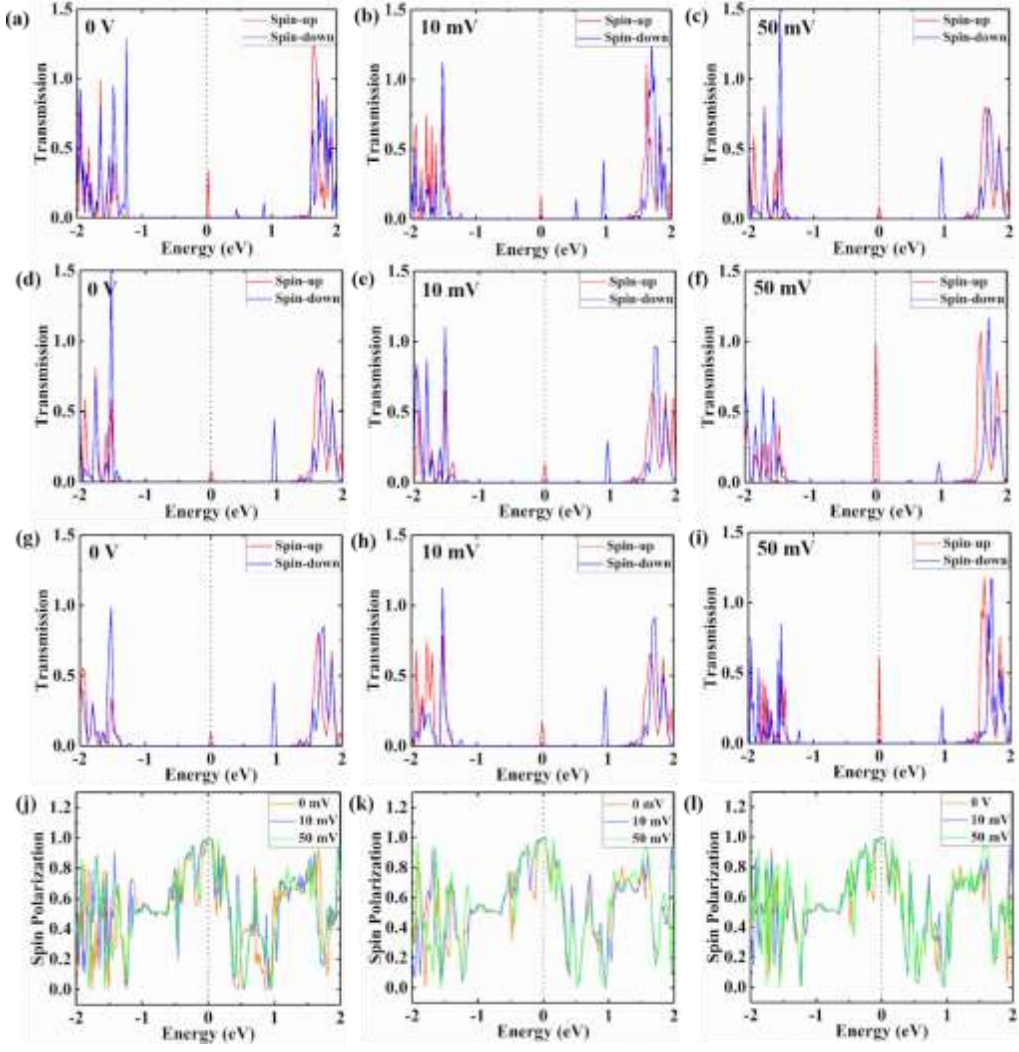

FIG. 4. Spin-dependent Transmission when Mn initial spin oriented along (a-c) a-direction, (d-f) b-direction, and (g-i) c-direction, at 0 mV, 10 mV, and 50 mV bias voltages. The calculated spin polarization at various bias voltages when the initial spin of Mn is oriented along (j) a-direction, (k) b-direction, and (l) c-direction.

The conductance is directly proportional to the transmission probability which indicates efficient electron transport. The observed transmission spectra with the initial spin of Mn atom oriented in a-, b-, and c-directions revealed distinct behavior of spin-up carriers exhibited transmission peaks precisely at the Fermi level, while spin-down carriers were located at an energy significantly away from the Fermi level. At zero bias voltage, $T_{up}$ [$T_{down}$] at the Fermi level were determined to be 0.338 [3.43 x $10^{-5}$], 0.0819 [4.35 x $10^{-5}$], and 0.098 [4.165 x $10^{-5}$] oriented along a-, b- and, c-directions, respectively. This substantial asymmetry confirmed a resonant behavior for $T_{up}$ and an anti-resonant peak for $T_{down}$ at the Fermi level. To quantitatively assess the asymmetry, we calculated the spin-filter efficiency (η) using the following formula below[51,52]

$$\eta = \frac{T_{up}(E_F) - T_{down}(E_F)}{T_{up}(E_F) + T_{down}(E_F)} \quad (6)$$

The absence of spin-down transmission at the Fermi level, resulting in a zero bias spin-filter efficiency (η) of approximately 99.9% in all the spin-oriented devices, unequivocally confirms the spin-selective nature of the Mn:H-SiQD channel. Moreover, in a-direction spin-oriented transmission spectra, we observed as the bias voltage was increased to 10 mV and 50 mV, the $T_{up}$ near the Fermi level exhibited a decrease, while the $T_{down}$ increased noticeably, particularly at energy 0.5 eV and 0.9 eV. In this, we have also observed the transmission of spin-down electrons through a nonbonding state (around 0.5 eV) at 0 mV and 10 mV. However, the opposite behavior is observed in b- and c-direction spin-oriented devices that on increasing the bias voltage $T_{up}$ increases and $T_{down}$ decreases. the spin-down level did not manifest as a peak at 0.5 eV in the c-direction spin-oriented device, whereas at 50 mV bias voltage, it appears in the b-direction spin-oriented device. Further, the calculated spin polarization using Eq. 4, illustrated in Fig. 4(J-l), indicated that the maximum spin polarization occurred precisely at the Fermi level for all bias voltages, reaching approximately 0.99. This value signifies the effective spin-filtering capability of the device. Remarkably, the calculated conductance values at Fermi level $G(E_F)$ at various bias voltages for all spin direction-oriented devices are listed in Table S2. As we know at the Fermi level the contribution of $T_{down}$ is negligible so these conductance values are due to the contribution of spin-up states. Here we found fractional conductance values which indicate the device does not behave as a perfect conductor but the conductance values are not very low, which indicates the advantage of Mn:H-SiQD for spintronic applications.

The thorough analysis developed a better understanding of the device's behavior under diverse conditions. Moreover, our study established a direct correlation between the locations of the spin levels in the DOS and the corresponding transmission peaks. This dependency indicates that there is no lead to lead direct tunneling and determines the significant role played by these specific orbitals in the complex process of electron transport. These findings align with the desirable properties required for a transmission channel to exhibit a strong spin-polarized transmission signal near the Fermi level.

In summary, our study involved a systematic investigation of 3d TM atoms doped at substitutional sites in H-SiQDs. Both PBE and HSE DFT functional approaches were employed for a comparative analysis. Our calculations on energetics and magnetic stability revealed that Mn and Fe are highly favorable dopants, displaying significant magnetic moments. Detailed DOS studies confirmed that Mn:H-SiQD exhibited superior stability when compared to Fe:H-SiQD. Furthermore, we demonstrated that in the quantum transport measurement, Mn:H-SiQD exhibits a spin filtering efficiency of 99.9% at the Fermi level with the initial spin orientated in a-, b-, and c-directions. This remarkable spin filtering efficiency is attributed to strong p-d hybridization between Si and Mn atoms within our nanostructure. We also investigated the impact of bias voltages on the spin filtering efficiency, which demonstrates a correlation between the initial spin direction and the direction of electric current flow, resulting in different transmission values. The results of high spin-filtering efficiency in the high-conductance regime could be useful for designing spin valves, magnetic tunnel junctions, and spin transistors.

See the supplementary information for computational details, the band structure of bare SiQD and Hydrogenated SiQD calculated by the HSE approach. Additionally, it provides details on the band structure of Mn:H-SiQD using both PBE and HSE approaches (Fig. S2 (a-d)), as well as Fe:H-SiQD using the HSE approach (Fig. S2(e and f)), with corresponding bandgap values listed in table S1. The Total and orbital decomposed DOS, Si atom orbital decomposed DOS, and partial density of states calculated by the HSE approach are illustrated in Fig. S3. Table S3 reveals the effects of various bias voltages on transmission values and conductance values at the Fermi level for different spin orientations in the Mn:H-SiQD spin-filtering device.

The authors express their gratitude to Prof. Sparsh Mittal for providing computational resources under Project No. ECR/2017/000622. Additionally, heartfelt thanks go to Prof. Sanjeev Manhas for his computational support under MIT-896-ECD EICT Academy IIT Roorkee. This endeavor received partial support from DST-SERB (Project No. ECR/2017/001050) and IIT Roorkee (Project No. FIG-100778-PHY), India. H.A. extends appreciation to the UGC, India, for the research scholarship.

DATA AVAILABILITY

The data that support the findings of this study are available within the article.

REFERENCES

[1] S. A. Wolf, D. D. Awschalom, R. A. Buhrman, J. M. Daughton, S. Von Molnár, M. L. Roukes, A.Y. Chtchelkanova, D.M. Treger, Science 294, 1488–1495 (2001).
[2] H. Ohno, D. Chiba, F. Matsukura, T. Omiya, E. Abe, T. Dietl, Y. Ohno, K. Ohtani, Nature. 408, 944–946 (2000).
[3] P. Recher, E. V. Sukhorukov, D. Loss, Phys. Rev. Lett. 85, 1962–1965 (2000).
[4] J. S. Moodera, T. S. Santos, T. Nagahama, J. Phys. Condens. Matter. 19, (2007).
[5] J. B. Moussy, J. Phys. D. Appl. Phys. 46, (2013).
[6] P. Li, C. Xia, Z. Zhu, Y. Wen, Q. Zhang, H.N. Alshareef, X.X. Zhang, Adv. Funct. Mater. 26, 5679–5689 (2016).
[7] R. A. Niyazov, D. N. Aristov, V. Y. Kachorovskii, Npj Comput. Mater. 6, 33–38 (2020).
[8] A. Saffarzadeh, G. Kirczenow, Phys. Rev. B. 104, 155406 (2021).
[9] A. Saffarzadeh and G. Kirczenow, Phys. Rev. B 102, 235420 (2020).
[10] A. Saffarzadeh and G. Kirczenow, Phys. Rev. B 106, 085416 (2022)
[11] M. Müller, M. Luysberg, and CM. Schneider, Appl. Phys. Lett. 98, 142503 (2011).
[12] M. Tanaka, M. Furuta, T. Ichikawa, M. Morishita, YM. Hung, S. Honda, T. Ono, and K. Mibu, Appl. Phys. Lett. 122, 042401 (2023).
[13] Y. Zheng, X. Ma, F. Yan, H. Lin, W. Zhu, Y. Ji, R. Wang, and K. Wang, npj 2D Materials and Applications 6(1), 62 (2022).
[14] F. Honglei, G. Shi, D. Yan, Y. Li, Y. Shi, Y. Xu, P. Xiong, and Y. Li, Appl. Phys. Lett. 121, 142402 (2022).
[15] J. Sun, R. S. Deacon,, X. Liu, J. Yao, and K. Ishibashi, Appl. Phys. Lett. 117, 052403 (2020).
[16] P. Recher, E.V. Sukhorukov, and D. Loss, Phys. Rev. Lett., 85(9), 1962-1965, (2000).
[17] H. Kiyama, T. Fujita, S. Teraoka, A. Oiwa, and S. Tarucha, Appl. Phys. Lett., 104(26), 263101 (2014).
[18] J. R. Hauptmann, J. Paaske, and P. E. Lindelof, Nature Physics, 4(5), 373-376 (2008).
[19] A. Bonanni, T. Dietl, Chem. Soc. Rev. 39, 528–539 (2010).
[20] T. Dietl, Nature materials 9(12), 965-974 (2010).
[21] D. Ferrand, J. Cibert, A. Wasiela, C. Bourgognon, S. Tatarenko, G. Fishman, T. Andrearczyk, J. Jaroszyński, S. Koleśnik, T. Dietl, B. Barbara, D. Dufeu, Phys. Rev. B 63, 085201 (2001).
[22] C. Liu, F. Yun, H. Morko c c, J. Mater. Sci. Mater. Electron. B. 6, 555–597 (2005).
[23] H. J. Choi, H. K. Seong, J. Chang, K. I. Lee, Y. J. Park, J. J. Kim, S. K. Lee, R. He, T. Kuykendall, P. Yang, Adv. Mater. 17, 1351–1356 (2005).
[24] M. Sawicki, D. Chiba, A. Korbecka, Y. Nishitani, J.A. Majewski, F. Matsukura, T. Dietl, H. Ohno, Nat. Phys. 6, 22–25 (2010).
[25] Y. D. Park, A. T. Hanbicki, S.C. Erwin, C. S. Hellberg, J. M. Sullivan, J. E. Mattson, T. F. Ambrose, A. Wilson, G. Spanos, and B. T. Jonker, Science 295(5555), 651-654 (2002).
[26] H. Dery, Y. Song, P. Li, and I. Zutic, Appl. Phys. Lett. 99, 082502 (2011).

[27] H. W. Wu, C. J. Tsai, and L. J. Chen, Appl. phys. lett. 90, 043121 (2007).

[28] P. R. Bandaru; J. Park; J. S. Lee; Y. J. Tang; L.-H. Chen; S. Jin; S. A. Song; J. R. O'Brien, Appl. phys. lett. 89, 112502 (2006).

[29] D. N. Jamieson, C. Yang, T. Hopf, S. M. Hearne, C. I. Pakes, S. Prawer, M. Mitic, E. Gauja, S. E. Andresen, F. E. Hudson, A. S. Dzurak, R. G. Clark, Appl. Phys. Lett. 86, 202101 (2005).

[30] M. Tabe, D. Moraru, M. Ligowski, M. Anwar, R. Jablonski, Y. Ono, and T. Mizuno, Phys. Rev. lett., 105(1), 016803 (2010).

[31] F. Abualnaja, W. He, K. L. Chu, A. Andreev, M. Jones, and Z. Durrani, Appl. Phys. Lett. 122, 233504 (2023).

[32] A. Samanta, M. Muruganathan, M. Hori, Y. Ono, H. Mizuta, M. Tabe, D. Moraru, Appl. Phys. Lett. 110, 093107 (2017).

[33] P. Yadav, H. Arora, and A. Samanta. Appl. Phys. Lett. 122, 083502 (2023).

[34] A. Samanta, D. Moraru, T. Mizuno, and M. Tabe, Scientific Reports 5(1), 17377 (2015).

[35] B. E. Kane, Nature **393**, 133 (1998).

[36] J. J. Pla, K. Y. Tan, J. P. Dehollain, W. H. Lim, J. J. Morton, D. N. Jamieson, A. S. Dzurak, and A. Morello, Nature, 489(7417), 541-545 (2012).

[37] M. M. Munia, S. Monir, E. N. Osika, M. Y. Simmons, and R. Rahman, Phys. Rev. Appl., 21(1), 014038 (2024).

[38] R. Jansen, Nature materials 11(5), 400-408 (2012).

[39] O. M. J. Van't Erve, A. L. Friedman, C. H. Li, J. T. Robinson, J. Connell, L. J. Lauhon, and B. T. Jonker, Nature communications 6(1), 7541 (2015).

[40] S. Zhang, S. A. Dayeh, Y. Li, S. A. Crooker, D. L. Smith, and S. T. Picraux, Nano letters, 13(2), 430-435 (2013).

[41] H. Arora, A. Samanta, Phys. Chem. Chem. Phys. 25, 2999–3010 (2022).

[42] K. Momma, F. Izumi, J. Appl. Crystallogr. 41, 653–658 (2008).

[43] S. Chinnathambi, S. Chen, S. Ganesan, N. Hanagata, Adv. Healthc. Mater. 3, 10–29 (2014).

[44] E. Durgun, N. Akman, S. Ciraci, Phys. Rev. B 78, 195116 (2008).

[45] M. V. Wolkin, J. Jorne, P.M. Fauchet, G. Allan, C. Delerue, Phys. Rev. Lett. 82, 197–200 (1999).

[46] Q. Xu, J. Li, S. S. Li, J. B. Xia, J. Appl. Phys. 104 (2008).

[47] H. A. Jahn and E. Teller, Proc. R. Soc. London. 161, 220–235 (1937).

[48] A. Kyono, S.A. Gramsch, T. Yamanaka, D. Ikuta, M. Ahart, B.O. Mysen, H. kwang Mao, R.J. Hemley, Phys. Chem. Miner. 39, 131–141 (2012).

[49] J. B. Goodenough, Journal of Physics and Chemistry of Solids 25(2), 151-160 (1964).

[50] L. Hu, J. Han, and G. Gao, Appl. Phys. Lett., 123, 052401 (2023).

[51] S. Aryal, R. Pati, Nanoscale Adv. 2, 1843–1849 (2020).

[52] T. Guo, X. Xu, H. Zhang, Y. Xie, H. Yang, X. Rui, Y. Sun, X. Yao, B. Wang, and X. Zhang, Appl. Phys. Lett. 124(6), 062404 (2024).

# Supporting Material

# Room Temperature Spin Filtering and Quantum Transport with Transition Metal-Doped Silicon Quantum Dot

Hemant Arora[1] and Arup Samanta*[1, 2]

[1]Quantum/Nano Science and Technology Laboratory, Department of Physics, Indian Institute of Technology Roorkee, Roorkee-247667, Uttarakhand, India

[2]Centre of Nanotechnology, Indian Institute of Technology Roorkee, Roorkee-247667, Uttarakhand, India

*Corresponding author: arup.samanta@ph.iitr.ac.in*

This supplementary material presents information on computational details, and band structure of b-SiQD and H-SiQD calculated by the HSE approach. Additionally, it provides details on the band structure of Mn:H-SiQD using both PBE and HSE approaches (Fig. S2 (a-d)), as well as Fe:H-SiQD using the HSE approach (Fig. S2(e and f)), with corresponding bandgap values listed in table S1. The orbital decomposed, atom decomposed, and partial density of states calculated by the HSE approach are illustrated in Fig. S3. Table S3 reveals the effects of various bias voltages on transmission values and conductance values at the Fermi level for different spin orientations in the Mn:H-SiQD spin-filtering device.

## Computational details

The first principle calculations were studied using a linear combination of atomic orbitals (LCAO) method as implemented in the QuantumATK software[1] within the framework of DFT.[2] In this study, the generalized gradient approximation (GGA) is used to represent electron exchange and interactions, whereas the Perdew, Burke, and Ernzerhof (PBE) technique describes electron-ion interactions.[3] The norm-conserving pseudopotential of the Troullier-Martins type[4] and an optimized double-polarized basis set are used to perform all computations for spin-unpolarized and spin-polarized structures. The energy cut-off was set to 45 Hartree along with the optimized k-point set[5] (2×2×2) to converge energy difference within the tolerance of 1 x $10^{-4}$ Hartree. Furthermore, using the LBGFS optimizer approach,[6] the geometry was relaxed and the atomic positions were relaxed until the maximum force was less than 0.001 eV/Å without utilizing any lattice parameter constraints. In the optimized geometry of H-SiQD, the Si-Si bond length is found to be 2.35 Å. The TM-doped H-SiQD structure optimized with the same parameters and observed bond lengths between TM-Si atoms along with the range of variation in tetrahedral angles are reported in Table 1. The spin-polarized generalized gradient approximation (SGGA) with PBE functional was used to obtain electrical and magnetic characteristics for TM:H-SiQD. Additionally, the Heyd-Scuseria-Ernzerhof (HSE) approach using the HSE06 hybrid functional is also used to obtain more accurate results. According to reports, for group IV semiconductors, the HSE approach is in good agreement with experimental results.[7]

Further, the electronic transport properties were investigated by constructing a device with TM:H-SiQD sandwiched between two gold electrodes. The electrode temperature is set to 300 K, assumed to represent room temperature operation. These transport properties are evaluated by using DFT with non-equilibrium Green's function (NEGF) formalism.[8] We employed the SGGA-PBE approach for transport calculations, but due to the limitation of the computational capacity HSE approach is not utilized. The spin-resolved bias-dependent transmission coefficients are obtained as

$$T_\sigma (E, V_b) = T_r[\Gamma_L G_r \Gamma_R G_a] (E, V_b) \qquad (1)$$

Where $G_r$ and $G_a$ are retarded and advanced Green's function, and $\Gamma_L$, $\Gamma_R$ are the self-energy matrices that describe the coupling of the central scattering region with the left and right side electrodes.

## Band structure

Fig. S1(a) and S1(b) depict the band structure calculations obtained using the HSE approach. The resulting band dispersion appears largely similar to that derived from the PBE approach, yet a slight shift in the bands is observable. Fig. S2(a-f) depicts the spin-up and spin-down band structures of TM:H-SiQD, the bands are nearly identical in the valence and conduction bands, while additional bands near the Fermi level suggest occupancy by extra electrons in hybrid states. The flatness of these bands indicates restricted electron mobility, crucial for the formation of localized magnetic moments.[9] Furthermore, the bandgap values obtained from spin-polarized band structures confirm the semiconducting nature. The calculated band structures using the HSE approach exhibit similar dispersion curves but with band shifts resulting in wider band gaps. Remarkably, Mn and Fe-doped H-SiQD exhibit splitting into three levels, as discussed in the article.

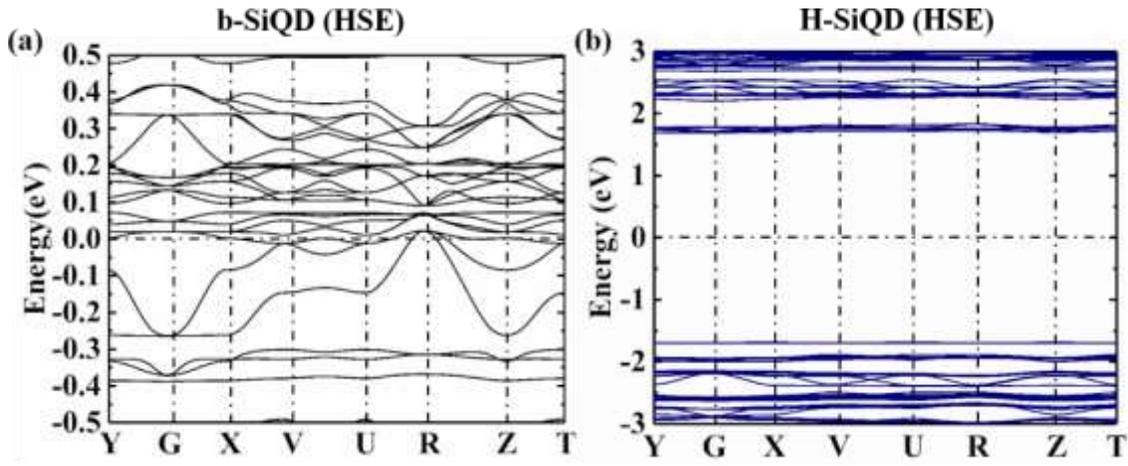

FIG. S1 The electronic band structures of (a) b-SiQD, (b) H-SiQD by HSE approach.

TABLE S1. The spin-up and spin-down bandgap values of Cr, Mn, and Fe doped H-SiQD by PBE and HSE approach.

| Structure | Approach | Bandgap (eV) Spin-up | Bandgap (eV) Spin-down |
|---|---|---|---|
| Mn:H-SiQD | PBE | 0.12 | 1.68 |
| | HSE | 0.6 | 2.87 |
| Fe:H-SiQD | HSE | 0.88 | 2.66 |

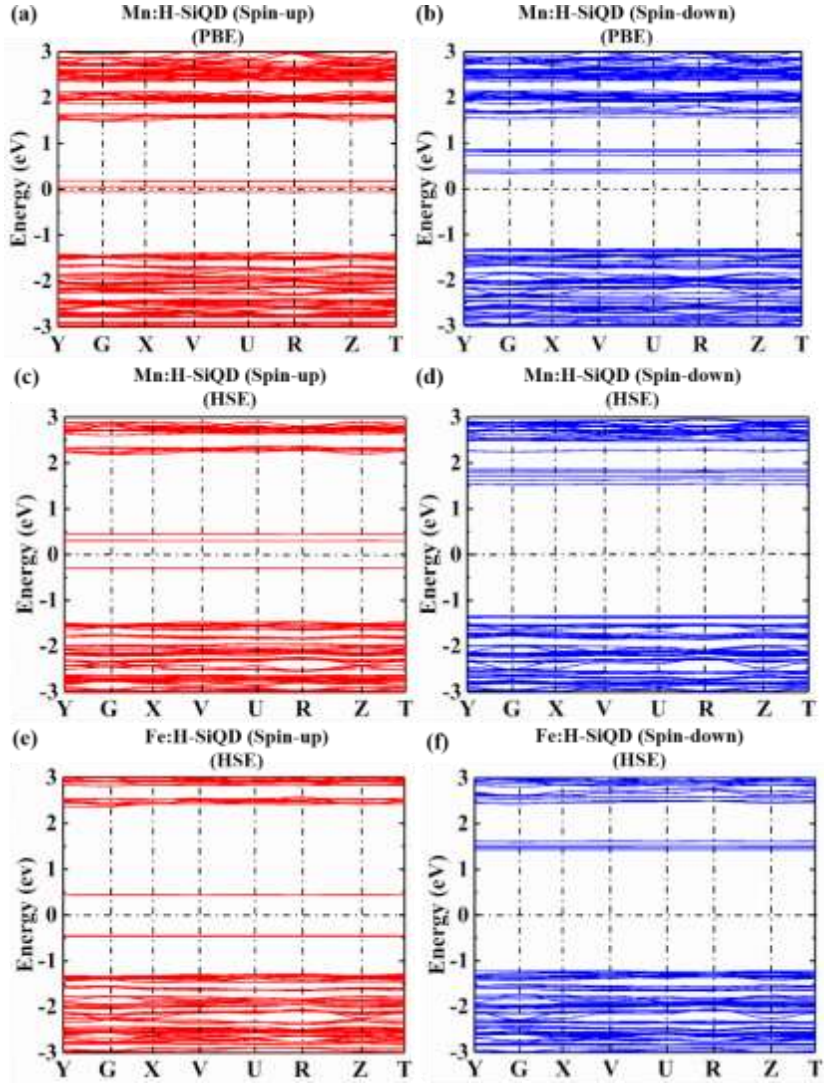

FIG. S2 The electronic band structure of Mn:H-SiQD (a) spin-up direction (b) spin-down direction calculated by PBE approach, Mn:H-SiQD (c) spin-up direction (d) spin-down direction, Fe:H-SiQD (e) spin-up direction (f) spin-down direction, calculated by HSE approach.

## Density of States

Fig. S3(a-f) showed atom decomposed, orbital decomposed density of states, and partial density of state for TM:H-SiQD by HSE approach. The fivefold degeneracy of TM atoms splits into three-fold $t_{2d}$ and two-fold $e_d$ states under tetrahedral crystal field splitting due to interactions with cation vacancy dangling bonds. These $t_{2d}$ states couple with $t_{2p}$ states, forming bonding and anti-bonding $t_2$ states, leading to partially filled spin-up and spin-down states near the Fermi level. This partial filling demonstrates strong p-d hybridization between Si 3p and TM 3d orbitals, evident in atom and orbital decomposed DOS of Si and Mn. Bonding and anti-bonding $t_2$ states are dominated by d orbital states (termed "crystal field resonance") or 3p orbital states (referred to as "dangling bond hybrids"). The schematic energy level diagram of the hybridization between the 3d level of TM and 3p level of cation vacancy is shown in Fig. S4.[10] The hybridization between TM 3d and 3p levels of cation vacancy results in bonding $t_2$ states primarily containing TM $t_2$ orbitals ($t_{2+}^{CFR}$ and $t_{2-}^{CFR}$) and anti-bonding $t_2$ states with dangling bond $t_2$ orbitals ($t_{2+}^{DBH}$ and $t_{2-}^{DBH}$). Additionally, TM e states create nonbonding levels ($e_{+}^{CFR}$ and $e_{-}^{CFR}$).

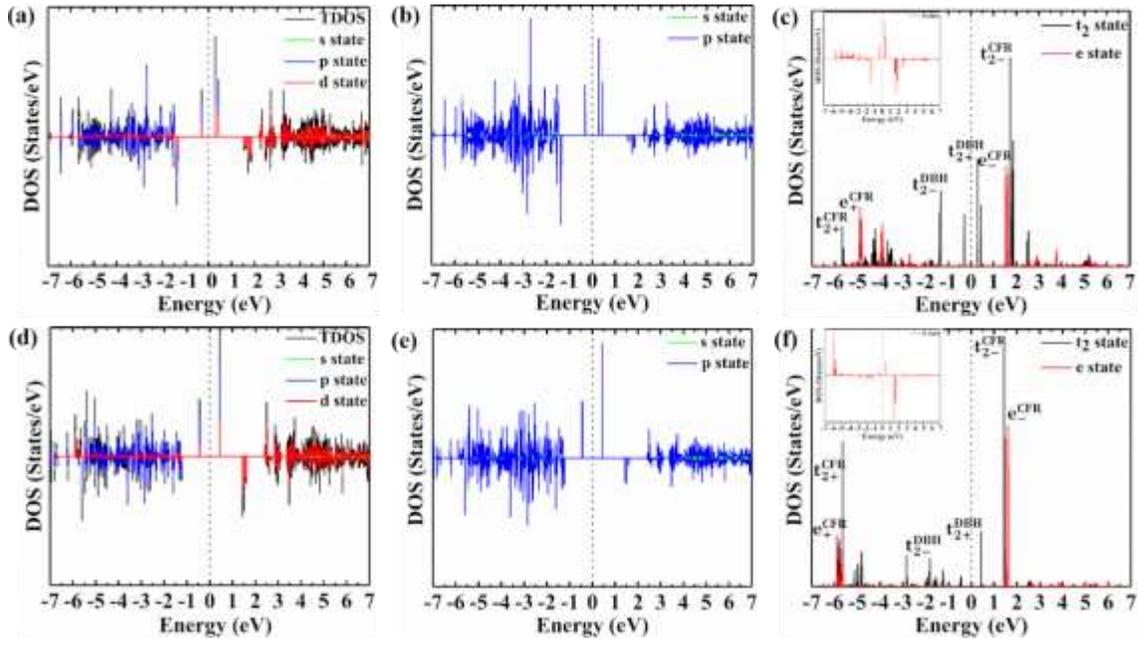

FIG. S3 The spin-up and spin-down (a) TDOS and atom decomposed DOS, (b) orbital decomposed DOS of Si atoms (c) Mn atom PDOS, of Mn:SiQD. The spin-up and spin-down (d) atom decomposed DOS, (e) orbital decomposed DOS of Si atoms (f) Fe atom PDOS, of Fe:SiQD. where the $t_2$ and e symmetry as well as spin (+ and -) have been indicated. The d orbital DOS of TM atoms are given in the inset.

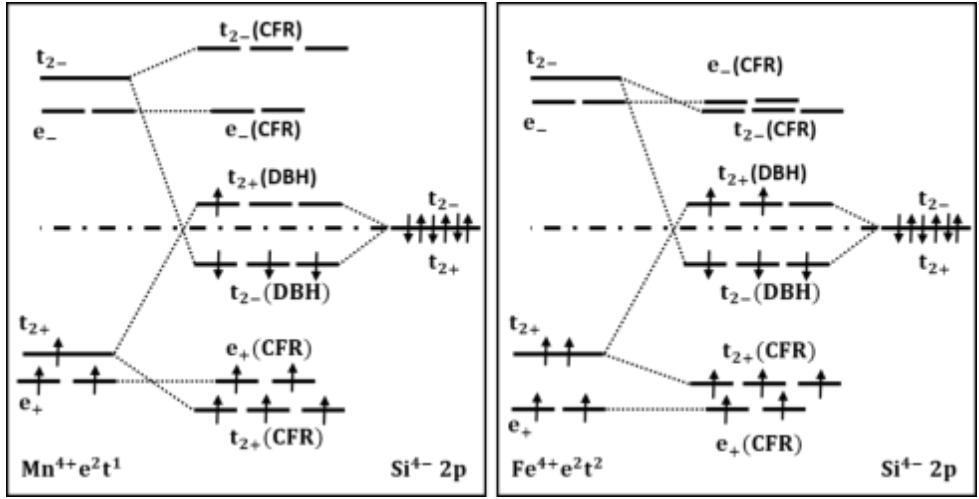

Fig. S4(a-b) The schematic energy level diagram of Mn, and Fe-doped H-SiQD, where the TM ion levels on the left side, dangling bond levels of Si on the right side, and the central region is showing the coupled levels generated from the interaction.

TABLE S2. The transmission and Conductance values at the Fermi level for different orientations of the initial spin of Mn atom at various bias voltages.

| Bias voltage (mV) | Transmission value for spin-up and spin-down state at Fermi level [$T_{up}$ ($T_{down}$)] | | | Conductance at Fermi level ($e^2/h$) | | |
|---|---|---|---|---|---|---|
| | a direction | b direction | c direction | a direction | b direction | c direction |
| 0 | 0.338 (3.43 x $10^{-5}$) | 0.0819 (4.35 x $10^{-5}$) | 0.098 (4.165 x $10^{-5}$) | 0.338 | 0.0819 | 0.098 |
| 10 | 0.169 (8.41 x $10^{-5}$) | 0.131 (5.75 x $10^{-5}$) | 0.169 (8.41 x $10^{-5}$) | 0.169 | 0.131 | 0.169 |

| 50 | 0.0819 (4.35 x 10⁻⁵) | 0.989 (3.65 x 10⁻⁵) | 0.609 (3.39 x 10⁻⁵) | 0.0819 | 0.989 | 0.609 |

**References**

[1] S. Smidstrup, T. Markussen, P. Vancraeyveld, J. Wellendorff, J. Schneider, T. Gunst, B. Verstichel, D. Stradi, P.A. Khomyakov, U.G. Vej-Hansen, M.E. Lee, S.T. Chill, F. Rasmussen, G. Penazzi, F. Corsetti, A. Ojanperä, K. Jensen, M.L.N. Palsgaard, U. Martinez, A. Blom, M. Brandbyge, K. Stokbro, J. Phys. Condens. Matter. 32 (2020).
[2] W.K. AND, L.J. SHAM, Phys. Rev. 140, A 1133 (1965).
[3] J.P. Perdew, K. Burke, M. Ernzerhof, Phys. Rev. Lett. 77, 3865–3868 (1996).
[4] N. Troullier, J.L. Martins, Phys. Rev. B. 43 1993–2006 (1991).
[5] H.J.M. and J.D. Pack, Phys. Rev. B. 13, 5188–5192 (1976).
[6] Dong C. Liu, Jorge Nocedal, Math. Program. 45 503–528 (1989).
[7] Y. Tan, M. Povolotskyi, T. Kubis, T.B. Boykin, G. Klimeck, Phys. Rev. B. 94, 045311 (2016).
[8] M. Brandbyge, J.L. Mozos, P. Ordejón, J. Taylor, K. Stokbro, Phys. Rev. B 65, 165401 (2002).
[9] Kusakabe, K. & Maruyama, M., *Phys. Rev. B - Condens. Matter Mater. Phys.* **67**, 924061–924064 (2003).
[10] Xu, Q., Li, J., Li, S. S. & Xia, J. B., *J. Appl. Phys.* **104**, (2008).